\documentclass[twocolumn,showpacs,aps,prl,superscriptaddress]{revtex4}

\usepackage{graphicx}
\usepackage{amsmath}

\begin{document}

\title{Microwave imaging of mesoscopic percolating network in a manganite thin film}

\author{Keji Lai}
\affiliation{Department of Applied Physics and Geballe Laboratory
for Advanced Materials, Stanford University, Stanford, CA 94305}
\author{Masao Nakamura}
\affiliation{Cross-Correlated Materials Group (CMRG), ASI, RIKEN,
Wako, 351-0198, Japan}
\author{Worasom Kundhikanjana}
\affiliation{Department of Applied Physics and Geballe Laboratory
for Advanced Materials, Stanford University, Stanford, CA 94305}
\author{Masashi Kawasaki}
\affiliation{Cross-Correlated Materials Group (CMRG), ASI, RIKEN,
Wako, 351-0198, Japan}
\affiliation{WPI-Advanced Institute for
Materials Research (AIMR), Tohoku University, Sendai, 980-8577,
Japan}
\author{Yoshinori Tokura}
\affiliation{Cross-Correlated Materials Group (CMRG), ASI, RIKEN,
Wako, 351-0198, Japan}
\affiliation{Department of Applied Physics,
University of Tokyo, Tokyo 113-8586, Japan}
\author{Michael A. Kelly}
\affiliation{Department of Applied Physics and Geballe Laboratory
for Advanced Materials, Stanford University, Stanford, CA 94305}
\author{Zhi-Xun Shen}
\affiliation{Department of Applied Physics and Geballe Laboratory
for Advanced Materials, Stanford University, Stanford, CA 94305}

\date{\today}

\begin{abstract}

Many unusual behaviors in complex oxides are deeply associated with
the spontaneous emergence of microscopic phase separation. Depending
on the underlying mechanism, the competing phases can form ordered
or random patterns at vastly different length scales. Using a
microwave impedance microscope, we observed an orientation-ordered
percolating network in strained Nd$_{0.5}$Sr$_{0.5}$MnO$_3$ thin
films with a large period of 100 nm. The filamentary metallic
domains align preferentially along certain crystal axes of the
substrate, suggesting the anisotropic elastic strain as the key
interaction in this system. The local impedance maps provide
microscopic electrical information of the hysteretic behavior in
strained thin film manganites, suggesting close connection between
the glassy order and the colossal magnetoresistance effects at low
temperatures.

\end{abstract}
\maketitle

Doped cuprate superconductors and colossal magnetoresistive (CMR)
manganites, the two most studied complex oxides, exhibit rich phase
diagrams as a result of the simultaneously active charge, spin,
orbital, and lattice degrees of freedom \cite{Dagotto,Tokura}.
Recent work on these strongly correlated materials has shown that
multiple states can coexist near certain phase boundaries, a
scenario known as microscopic phase separation \cite{Dagotto}. The
configurations of these spatially inhomogeneous phases reflect the
underlying interactions. When the long-range Coulomb interaction
prevails, the competing phases usually form nanometer-scale orders,
because of the electrostatic energy penalty for macroscopic phase
separation \cite{Yunoki,Vershinin,Hanaguri,Howald}. For
self-organized patterns at larger length scales, weaker long-range
interactions, such as the elastic strain arising from either the
cooperative lattice distortions or lattice mismatch between
substrates and epitaxial thin films, become the dominant factors
\cite{Mathur,Ahn,Burgy}. Finally, the unavoidable quenched disorders
in real materials always introduce short-range potential
fluctuations, which usually smear out the orders or even result in
micrometer-sized clusters with random shapes \cite{Moreo}.

Many physical properties affected by the phase separation, such as
the local density of states
\cite{Vershinin,Hanaguri,Howald,Fath,Renner}, the local
magnetization \cite{Zhang,Wu,Loudon}, and the atomic displacement
\cite{Uehara}, can be spatially mapped out by established microscopy
tools. For CMR manganites with drastic resistance changes at
different temperatures (T) and magnetic fields (H), the local
resistivity ($\rho$) has a large span that makes spatially resolved
DC measurements challenging. Imaging with high-frequency AC-coupled
local probes is thus desirable. We carried out a microwave impedance
microscopy (MIM) study \cite{Lai1,Lai2} on manganite thin films.
Unlike other GHz microscopes \cite{Rosner}, the cantilever probe is
well shielded to reduce the stray fields \cite{Lai2}. In the
microwave electronics, the high-Q resonator \cite{Wang} susceptible
to environmental conditions is eliminated so that the system can be
implemented under variable temperatures (2-300K) and high magnetic
fields (9T). Our cryogenic MIM is therefore a versatile tool to
investigate various electronic phase transitions. Only the imaginary
part of the tip-sample impedance is presented here as the local
resistivity information is fully captured by the capacitive channel
(MIM-C).

\begin{figure}[!t]
\begin{center}
\includegraphics[width=3.2in]{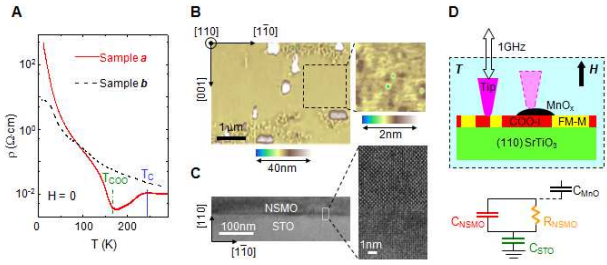}
\end{center}
\caption{\label{1} (A) Zero-field  (T) curves of two samples $a$
(30nm NSMO/STO) and $b$ (60nm NSMO/STO). The metallic temperature
region between the bulk T$_C$ (blue) and T$_{COO}$ (green) is
present in sample $a$ but missing in sample $b$. At low-T,
$\rho_a$(T) diverges and $\rho_b$(T) saturates. (B) The AFM surface
topography contains MnO$_x$ precipitates ($\sim$20nm in height) and
corrugations. Crystal axes of the STO substrate are indicated. The
close-up shows an atomically-flat region 1$\sim$2$\mu$m in size.
Note the different false-color scales between the main figure (40nm)
and the close-up (2nm). (C) Cross-sectional TEM image of the
NSMO-STO interface of sample $a$. The good crystalinity and coherent
epitaxy are confirmed by a close-up high-resolution TEM picture. (D)
Schematics of the system setup (top) and the corresponding
lumped-element circuit (bottom). When the tip scans over the FM-M
domains (yellow), the small R$_{NSMO}$ shunts the capacitor
C$_{NSMO}$, giving a larger signal than that of the COO-I background
(red). The insulating particles (black), on the other hand,
introduce a series capacitor (C$_{MnO}$), resulting in lower MIM-C
signals.}
\end{figure}

We study Nd$_{0.5}$Sr$_{0.5}$MnO$_3$ (NSMO) thin films grown on
(110) SrTiO$_3$ (STO) substrates by pulsed-laser deposition. In
single crystal NSMO, both resistivity and magnetization measurements
show a paramagnetic (PM) to ferromagnetic (FM) transition at the
Curie temperature T$_C$ $\sim$ 250K and a charge/orbital-order (COO)
transition at T$_{COO}$ $\sim$ 160K \cite{Kuwahara}. When a magnetic
field is turned on at temperatures below T$_{COO}$, a dramatic
first-order phase transition from the antiferromagnetic COO
insulating (COO-I) state to the FM metallic (FM-M) state is
observed. We emphasize that two types of CMR, the T-driven PM-FM CMR
and the low-T H-driven CMR, are widely discussed in the literature
\cite{Aliaga} and we only focus on the latter in this work. Recent
effort in this model system has been devoted to epitaxial films
\cite{Nakamura,Wakabayashi}. To date, bulk-like behaviors are only
seen for films grown on (110) STO substrates, presumably due to the
strong dependence on the lattice strain. Even on the same substrate,
sample degradation resulting from partial loss of crystalinity and
epitaxial coherency is sometimes detected. We present data from two
samples, sample $a$ in which clear signatures of T$_C$ and T$_{COO}$
are present (Fig. 1A), and, for comparison, a degraded sample $b$,
which lacks metallic temperature regions. The samples were
characterized by atomic force microscope (AFM, Fig. 1B) and
cross-sectional transmission electron microscope (TEM, Fig. 1C). The
surface contains micrometer-sized precipitates, most likely MnO$_x$
\cite{Higuchi}, and some corrugations next to these particles. In
between the defective regions, there exist atomically flat areas
where pure electrical signals can be obtained. The contrast in MIM-C
images is qualitatively understood by the lumped-element circuit in
Fig. 1D.

\begin{figure}[!t]
\begin{center}
\includegraphics[width=3.2in]{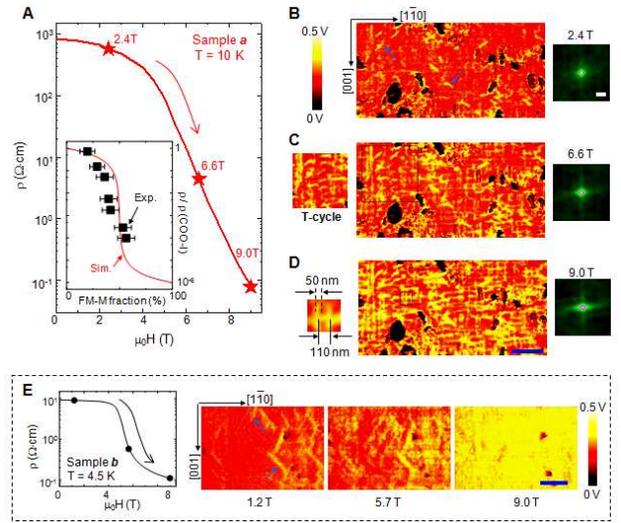}
\end{center}
\caption{\label{2} (A) $\rho_a$(H) during the field sweep at T =
10K. The three fields at which MIM images in (B-D) were taken are
labeled by the red stars. Inset, comparison between the 2D
square-lattice simulation ($\rho_{COO-I}\sim10^3\Omega$.cm and
$\rho_{FM-M}\sim10^{-3}\Omega$.cm) and the experimental data. The
large error bars and deviations near the percolation transition may
result from the surface defects, the finite resolution, and the fact
that percolation in NSMO is 3D in nature. (B-D) Microwave images
(scale bar 1$\mu$m) taken at 10K and under $\mu_0$H = 2.4T, 6.6T,
and 9.0T, respectively. MnO$_x$ particles appear black in the
images. At low fields, some isolated rod-like FM-M domains (yellow)
are indicated by blue arrows. Left inset of (C), measurement taken
at 15K and 6.5T after a thermal cycle. Left inset of (D), close-up
of a small region in (D) (black square) shows the typical width and
spacing of the filaments. The center portions of auto-correlation
image (scale bar 0.2$\mu$m) are shown on the right of (B-D), with a
notable center-cross seen at 9.0T. (E) Transport and three MIM
images (scale bar 1$\mu$m) of sample b taken at 4.5K. The low-field
FM-M rods are indicated by arrows.}
\end{figure}

Simultaneously taken transport data and low-T microwave images are
shown in Figs. 2A-D for sample $a$ and Fig. 2E for sample $b$. Both
the DC voltage and microwave excitation were kept low to avoid any
extrinsic perturbation. The insulating MnO$_x$ particles behaved as
field-independent markers. At low fields where $\rho$ barely
decreases, some rod-like FM-M domains (indicated by blue arrows)
tilted with respect to the [001] direction were observed on the
COO-I background in both samples. The presence of such low-field
conducting domains implies that phase separation already occurs
before the H-field is turned on. We note that sample $b$ showed
larger low-field FM-M areal fraction than sample $a$. As H was
increased to 6$\sim$7T, metallic areas grew from the low-field
nucleation sites; their positions and shapes indicate a certain
memory effect \cite{Sarma}. In a subsequent experiment, sample $a$
was warmed up to 250K and cooled back to 15K at H = 0. A field of
6.5T was turned on after this thermal cycle. Most FM-M domains, as
shown in the left inset of Fig. 2C, reappeared in the same locations
compared to the corresponding area in the main Fig. 2C (dotted box),
indicative of pinning by an intrinsic long-range energy landscape
and short-range disorder potentials. A striking distinction between
the two samples at intermediate fields is that the H-induced FM-M
filaments of sample $a$ (Fig. 2C) show clear directional ordering
and preferentially align along [001] and [1-10] axes of the
substrate, whereas no such feature is seen in sample b within our
spatial resolution. At 8$\sim$9T, the prominent FM-M filaments in
sample a form an interconnected percolating network. While the
smallest measured feature width (30-50nm) may be set by our spatial
resolution, the typical spacing $\sim$100nm is clearly resolved
here. As $\rho_a$(H) does not show any sign of saturation at 9T, the
FM-M domains should further expand at higher fields. $\rho_b$(H), on
the other hand, levels off at 9T, consistent with the nearly full
coverage of FM-M regions.

The salient liquid-crystal-like metallic network accompanying the
low-T CMR effect in sample $a$ is not seen by structural
characterization and must be electronic in origin. We can obtain
some insight by comparing this result with experiments on other
manganites. Nanometer-sized domains with no preferred directions
were observed in La$_{1-x}$Ca$_x$MnO$_3$ \cite{Fath,Tao} due to
Coulomb interaction. Large sub-micrometer clusters with no
particular shapes were imaged in (La,Pr)$_{1-x}$Ca$_x$MnO$_3$
\cite{Uehara} owing to the strong disorders. In the NSMO films, the
elastic strain must play a vital role since neither Coulomb
interaction nor quenched disorders can generate such mesoscopic
ordered network \cite{Dagotto,Tokura,Yunoki}. It is likely that the
accommodation strain \cite{Mathur,Wu} between the pseudo-cubic FM-M
and distorted orthorhombic COO-I phases is responsible for the
low-field FM-M rods, which appear in both sample $a$ and $b$. The
more important epitaxial strain imposed by the (110) STO substrate
\cite{Aliaga,Nakamura}, which is crucial for producing the bulk-like
behaviors, must account for the observed glassy orders in sample
$a$. The lack of a universal form of phase separation in both the
PM-FM \cite{Fath,Tao} and the low-T CMR effects \cite{Uehara} points
to the robustness of the phenomenon - the effect occurs because of
energetically competing states, while the exact CMR magnitude
depends on detailed microscopic configurations.

Taking the areal fraction of FM-M states as the probability of
connected bonds, we can compare the experimental data to the
square-lattice random resistor network simulation
\cite{Kirkpatrick,Mayr}. Despite some deviations near the threshold,
the agreement between experiment and modeling in the inset of Fig.
2A shows that the low-T CMR is indeed percolative in nature. Second,
the auto-correlation analysis is performed for Figs. 2B-D. The
nearly circular auto-correlation peak at 2.4T evolves into a clear
center-cross at 9T with a characteristic length $\sim$0.5$\mu$m. In
analogy to the stripes or checkerboard patterns seen in cuprates
\cite{Vershinin,Hanaguri,Howald}, the 100nm period may be set by the
long-range strain field, whereas the typical length of the filaments
is determined by the strength of disorders. Interestingly,
discernible anisotropy is seen in that the nematic domains along the
[1-10] direction are statistically more favorable than the [001]
direction, in fair agreement with the transport anisotropy observed
in control samples. In the strained NSMO/(110)STO films, the
in-plane lattice constant is locked to that of the substrate along
the [001] direction while relaxed along the [1-10] direction
\cite{Aliaga,Nakamura}. At the same time, the charge-ordered planes
are parallel to the (100) or (010) planes. As the low-T CMR effect
is accompanied by lattice deformation \cite{Kuwahara}, the metallic
domains may tend to expand along the more strain-free axis,
resulting in the in-plane anisotropy. The nematic phase at this
length scale provides a contrasting framework to understand the
stripe phenomenon that also breaks in-plane C$_2$-symmetry at much
shorter length scales \cite{Robertson}.

\begin{figure}[!t]
\begin{center}
\includegraphics[width=3in]{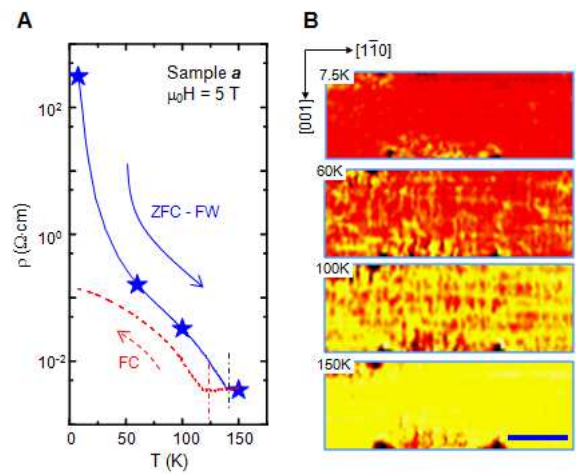}
\end{center}
\caption{\label{3} (A) $\rho_a$(T) at $\mu_0$H = 5T with the
temperatures at (B) labeled. Before this experiment, sample $a$ was
cooled to 7.5K at H=0 followed by field sweep to 5T. The subsequent
field-cool data (red dashed line) are also included. The up-turn
temperatures (T$_{COO}$) are marked as dash-dot lines. (B) MIM
images (scale bar 1$\mu$m) taken by a second probe at 7.5K, 60K,
100K, and 150K from top to bottom, showing more FM-M domains at
different temperatures. Due to the change of impedance match at
different T, the full false-color scale is adjusted as 0.6V (7.5K),
0.7V (60K), 1.1V (100K), and 1.2V (150K), respectively.}
\end{figure}

The physical picture depicted above is further corroborated by
results at elevated temperatures, as shown in Fig. 3. Before
imaging, sample $a$ was prepared by zero-field-cool (ZFC) to 7.5K
and a field sweep to 5T. The field-warm (FW) resistivity data are
shown in Fig. 3A. From 7.5K to 60K, considerable expansion of FM-M
domains is seen, in conjunction with the resistivity drop by three
orders of magnitude due to both thermal activation and percolation.
At 100K, notable mesh-like FM-M network appears in the impedance
map. For temperatures above T$_{COO}\sim$140K at 5T, the COO-I
phases completely melt into metals, leaving only the insulating
particles in the otherwise uniform sample.

\begin{figure}[!t]
\begin{center}
\includegraphics[width=3.2in]{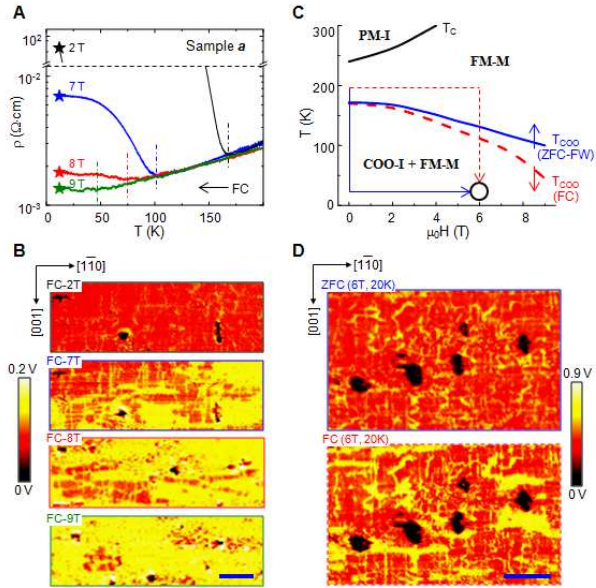}
\end{center}
\caption{\label{4} (A) Field-cool resistivity and (B) MIM images
(scale bar 1$\mu$m) at 2T, 7T, 8T, and 9T, respectively. T$_{COO}$
is denoted by vertical dash-dot lines. (C) Phase diagram of sample
$a$, showing the paramagnetic-insulator (PM-I), ferromagnetic-metal
(FM-M) and phase coexistence (COO-I + FM-M) regions demarcated by
transport signatures T$_C$ or T$_{COO}$ from ZFC-FW (field-warm,
solid blue line) and FC (dashed red line) processes. Two paths from
200K, 0T to 20K, 6T are sketched in the plot. (D) MIM images (scale
bar 1$\mu$m) taken at the same T = 20K and $\mu_0$H = 6T. The FC
image (bottom) contains much bigger conducting domains than the ZFC
one (top).}
\end{figure}

In Fig. 3A, $\rho_a$(T) during the cooling at 5T is also included,
showing the marked hysteretic behavior widely observed in
manganites. Using such field-cool (FC) process, one can access
states with much lower than the ZFC process. Figs. 4A and 4B show
the FC curves at four different fields and the corresponding
microwave images taken at 12K, which is below T$_{COO}$ for all
fields. As H increases, the continuous COO-I phases at FC-2T break
into isolated micrometer-sized domains (FC-7T), which continue to be
percolated through by FM-M filaments (FC-8T) and shrink down to
small droplets at FC-9T. Taking the transport signatures T$_C$ and
T$_{COO}$, we construct the phase diagram of this NSMO/STO sample in
Fig. 4C, where phase coexistence is denoted below T$_{COO}$. The
difference between the ZFC-FW and FC processes is shown by two lines
separating the FM-M and mixed phase regions. This phase diagram is
reminiscent of the one for single crystal NSMO \cite{Kuwahara}
except that the reentrant behavior reported there is beyond our
field range. Using MIM, the microscopic origin of the hysteresis can
be directly studied. In Fig. 4C, two paths arriving at the same
external conditions are shown - the ZFC process from 200K to 20K
followed by a field sweep to 6T, or field sweep to 6T at 200K before
FC to 20K. The two MIM images in Fig. 4D display remarkably
different percolating network. For the high-$\rho$ (1.4$\Omega$.cm)
ZFC state, glassy FM-M filaments are again observed in the COO-I
background. For the low-$\rho$ (0.02$\Omega$.cm) FC state, on the
other hand, the FM-M phases occupy a much larger portion and even
form micrometer-sized puddles elongated in the [1-10] direction.
While hysteresis during the low-T CMR effect is known in single
crystal NSMO from bulk measurements, tools like MIM enable real
space electrical imaging and demonstrate the strong dependence of
phase separation on local disorders and strain fields near the
multi-phase boundary.


\begin{thebibliography}{90}

\bibitem{Dagotto} E. Dagotto, Science {\bf 309}, 257 (2005).
\bibitem{Tokura} Y. Tokura, Rep. Prog. Phys. {\bf 69}, 797 (2006).
\bibitem{Yunoki} S. Yunoki et al., Phys. Rev. Lett. {\bf 80}, 845 (1998).
\bibitem{Vershinin} M. Vershinin et al., Science {\bf 303}, 1995
(2004).
\bibitem{Hanaguri} T. Hanaguri et al., Nature {\bf 430}, 1001
(2004).
\bibitem{Howald} C. Howald, H. Eisaki, N. Kaneko, M. Greven, and A.
Kapitulnik, Phys. Rev. B {\bf 67}, 014533 (2003).
\bibitem{Mathur}  N. D. Mathur and P. B. Littlewood, Solid State
Commun. {\bf 119}, 271 (2001).
\bibitem{Ahn} K. H. Ahn, T. Lookman, and A. R.
Bishop, Nature {\bf 428}, 401 (2004).
\bibitem{Burgy} J. Burgy, A. Moreo, and E.
Dagotto, Phys. Rev. Lett. {\bf 92}, 097202 (2004).
\bibitem{Moreo} A. Moreo, M. Mayr, A. Feiguin, S. Yunoki,
and E. Dagotto, Phys. Rev. Lett. {\bf 84}, 5568 (2000).
\bibitem{Fath} M. Fath et al., Science {\bf 285}, 1540 (1999).
\bibitem{Renner} Ch. Renner, G. Aeppli, B.-G. Kim, Yeong-Ah Soh, and S.-W.
Cheong, Nature {\bf 416}, 518 (2002).
\bibitem{Zhang} L. Zhang, C. Israel, A. Biswas, R. L. Greene, and
A. de Lozanne, Science {\bf 298}, 805 (2002).
\bibitem{Wu} W. Wu et al., Nat. Mat. {\bf 5}, 881 (2006).
\bibitem{Loudon} J. C. Loudon, N. D. Mathur and P. A. Midgley,
Nature {\bf 420}, 797 (2002).
\bibitem{Uehara} M. Uehara, S. Mori, C. H. Chen,
and S.-W. Cheong, Nature {\bf 399}, 560 (1999).
\bibitem{Lai1} K. Lai, W. Kundhikanjana, M. Kelly, and Z. X. Shen,
Rev. Sci. Instrum. {\bf 79}, 063703 (2008).
\bibitem{Lai2} K. Lai, W. Kundhikanjana, M. Kelly, and Z. X. Shen,
Appl.Phy.Lett. {\bf 93}, 123105 (2008).
\bibitem{Rosner} B. T. Rosner and D. W. van der Weide,
Rev. Sci. Instrum. {\bf 73}, 2505 (2002).
\bibitem{Wang} Z. Wang et al., J. Appl. Phys. {\bf 92}, 808 (2002).
\bibitem{Kuwahara} H. Kuwahara, Y. Tomioka, A. Asamitsu, Y. Moritomo, and Y.
Tokura, Science {\bf 270}, 961 (1995).
\bibitem{Aliaga} H. Aliaga et al., Phys. Rev. B {\bf 68}, 104405 (2003).
\bibitem{Nakamura} M. Nakamura, Y. Ogimoto, H. Tamaru, M. Izumi, and K. Miyano,
Appl. Phys. Lett. {\bf 86}, 182504 (2005).
\bibitem{Wakabayashi} Y. Wakabayashi et al., Phys. Rev. Lett. {\bf 96},
017202 (2006).
\bibitem{Higuchi} T. Higuchi et al., Appl. Phys. Lett. {\bf 95}, 043112 (2009).
\bibitem{Sarma} D. D. Sarma et al., Phys. Rev. Lett.
{\bf 93}, 097202 (2004).
\bibitem{Tao} J. Tao et al., Phys. Rev. Lett. {\bf 103}, 097202 (2009).
\bibitem{Kirkpatrick} S. Kirkpatrick, Rev. Mod. Phys.
{\bf 45}, 574 (1973).
\bibitem{Mayr} M. Mayr et al., Phys. Rev. Lett. {\bf 86}, 135 (2001).
\bibitem{Robertson} J. A. Robertson, S. A. Kivelson, E. Fradkin, A. C. Fang, and A.
Kapitulnik, Phys. Rev. B {\bf 74}, 134507 (2006).

\end{thebibliography}
\end{document}